\documentclass[12pt]{iopart}

\usepackage{graphicx}
\usepackage[usenames]{color}

\usepackage{iopams}  
\usepackage[colorlinks,linkcolor=red,anchorcolor=blue,
citecolor=blue]{hyperref}

\begin{document}

\title [Matter-waves in BECs with SO and Rabi couplings] 
{Matter-waves in Bose-Einstein condensates
with spin-orbit and Rabi couplings}

\author{Emerson Chiquillo}
\address{Escuela de F\'isica, Universidad Pedag\'ogica
y Tecnol\'ogica de Colombia (UPTC),\\
Avenida Central del Norte, Tunja, Colombia.}
\ead{emerson.chiquillo@uptc.edu.co}

\begin{abstract} 
We investigate the one-dimensional (1D) and two-dimensional
(2D) reduction of a quantum field theory starting from the
three-dimensional (3D) many-body Hamiltonian of 
interacting bosons with spin-orbit (SO) and Rabi couplings. 
We obtain the effective time-dependent 1D and 2D
nonpolynomial Heisenberg equations for both repulsive
and attractive signs of the inter-atomic interaction.
Our findings show that in case in which the many-body state
coincides with the Glauber coherent state, 
the 1D and 2D Heisenberg equations 
become 1D and 2D nonpolynomial Schr\"odinger equations
(NPSEs).
These models were derived in a mean-field approximation
from 3D Gross-Pitaevskii equation (GPE), describing a Bose-Einstein condensate (BEC) with SO and 
Rabi couplings. In the present work
self-repulsive and self-attractive localized solutions
of the 1D NPSE and the 1D GPE are obtained in a 
numerical form.
The combined action of SO and Rabi couplings
produces conspicuous sidelobes on the density profile,
for both signs of the interaction.
In the case of the attractive nonlinearity, an essential 
result is the possibility of getting an unstable
condensate by increasing of SO coupling.
\end{abstract}


\maketitle
 
\section{Introduction}

In last two decades, the ultracold atomic gases
have provided an important environment for studying 
quantum many-particle systems from the experimental
realization and the theoretical viewpoint.
In last few years, the huge interest in the ultracold 
gases field has allowed the experimental realization
of an artificial spin-orbit (SO) coupling in a neutral 
atomic Bose-Einstein condensate (BEC) by means of 
counterpropagating laser beams which couple
two atomic spin states \cite{SOC-1,soc-1}.
The SO coupling has been also created in 
fermionic atomic gases \cite{SOC-2,SOC-3}.
These experimental breakthroughs have led to significant
theoretical works, opening the door to a fascinating and
fast developing on SO coupled cold atoms research.

Basic theoretical and experimental aspects on SO coupled
degenerate atomic gases are introduced in the reviews
\cite{spin-orbit, spin-orbit-1}.
The theoretical activities have been devoted to 
the SO coupled Bose-Einstein condensates (BECs)
among others, including
vortex in rotating SO coupled BECs
\cite{vort-1,vort-2,vort-3,vort-4}.
Trapped two-dimensional (2D) 
atomic BECs with spin-independent interactions 
in the presence of isotropic SO coupling \cite{Santos-1,
vortex-5}.
In a two-component Bose gas confined in a 2D harmonic
oscillator (HO) potential, with an isotropic SO coupling
Rashba type, the ground state has a half-quantum angular 
momentum vortex configuration \cite{half-vortex}.
The ground-state properties of a weakly trapped 
spin-1 BEC with SO coupling were studied by
means of numerical and analytical methods in an external
Zeeman field \cite{spin1}.
From numerical simulation spin-$1/2$ and spin-1 BECs with
Rashba SO coupling were studied
\cite{Spinor-BEC}, identifying two different phases.
In one phase, the ground state is a single plane wave.
In the other phase, the condensate wave 
function is a standing wave, and it forms a spin stripe.
In  \cite{SOC-BECs}, low-energy stationary states of 
pseudospin-1 BECs in the presence 
of Rashba-Dresselhaus-type SO coupling were 
numerically investigated.
The quantum tricriticality and phase transitions in 
SO coupled BECs were studied by considering a 
SO coupled configuration of spin-1/2 interacting bosons
with equal Rashba and Dresselhaus couplings
in a mean-field approximation 
\cite{Tricriticality}.
In \cite{order-disorder} were studied isotropically 
interacting bosons with Rashba SO coupling. The
bosons condenses into a single momentum state of the 
Rashba spectrum via the mechanism of order by disorder.
The existence of antiferromagnetically ordered (striped) 
ground states in a 1D SO coupled system 
with repulsive atomic interactions
under the presence of a trapping HO it was demonstrated in
\cite{Malomed-1}. 
An experimental scheme to create SO coupling in spin-3
Cr atoms using Raman processes was proposed,
and the ground-state structures of a SO coupled Cr
condensate were studied \cite{SOC-experim}.
Thus showing, in addition to the stripe structures induced by 
the SO coupling, the magnetic dipole-dipole interaction 
gives rise to the vortex phase, in which a spontaneous spin 
vortex is formed. 
Solitons in a BEC with SO coupling were introduced in 
1D geometries
\cite{Bright-1,Bright-2,Bright-3,Bright-4} and 2D geometries
\cite{2D-1,2D-2,Salasnich-2D}.
Vortex-lattice solutions to the coupled mean-field 
equations with the SO coupling and optical lattice (OL)
potential were reported in \cite{Vortex-dynamics}. 
Vortex dynamics in SO coupled BECs was 
studied in \cite{Fetter,Fetter-2}.
In \cite{Adhikari-1} it was carried out a study 
by means of a variational approximation and
numerical solutions of localization of a noninteracting 
and weakly interacting SO coupled BEC in a quasiperiodic
bichromatic OL potential, confirming the
existence of stationary localized states in the presence 
of the SO and Rabi couplings for equal numbers of atoms in
the two components. The stability of plane waves in a 2D
SO coupled BEC was studied analytically in
\cite{stability-1, stability-2}.
Recently, has been studied the formation of bound states and
three-component bright vector solitons in a quasi-1D SO 
coupled hyperfine spin $f = 1$ BEC and the five component 
vector solitons in a quasi-1D SO coupled hyperfine spin-2
BEC using a numerical solution and a variational
approximation of a mean-field model 
\cite{Adhikari-2, Adhikari-3}.
It was considered a theory of collapse in BECs with
inter-atomic attraction for two different realizations 
of SO coupling \cite{Collapse-SOC-BEC}, 
the axial Rashba coupling and the balanced effectively
1D Rashba-Dresselhaus coupling.
The spin-dependent anomalous velocity in Rashba coupled 
BECs forms a centrifugal component in the density
flux opposite to that arising due to the attraction 
between particles and prevents the collapse at
a sufficiently strong coupling.
In BECs with balanced Rashba-Dresselhaus coupling, 
the spin-dependent velocity can spatially split the initial
state in 1D and form spin-projected wavepackets,
reducing the total condensate density. 
The quantum dynamics of a SO coupled BEC in a double-well 
potential was investigated in \cite{Josephson-1}. It was 
found that the SO coupling can significantly enhance the 
atomic interwell tunneling. 
By employing  the two-mode approximation from mean-field GPE
and numerical methods, the Josephson oscillations of
SO coupled BECs were studied in \cite{Josephson-2}.
This was carried out by analyzing the interplay between 
the inter-atomic interactions and the SO coupling and
the self-trapped dynamics of the inter-species imbalance. 

In the ultracold bosonic gases context at zero temperature,
a good theoretical tool for researching on dynamics of
dilute BECs is the time-dependent mean-field 3D GPE
\cite{Dalfovo}.
An interesting theoretical problem is the derivation
of reduced 1D and 2D models to studying the behavior of 
these systems. 
Experimentally these effective models are an 
ideal platform for testing many-body phenomena 
\cite{Bright-solitons1,Bright-solitons2,Giamarchi,
cazalilla2011}. 
Theoretically they have been adopted various approaches
to derive effective 1D and 2D reduced equations 
from the 3D GPE 
\cite{Muñoz-Delgado, Salasnich-NPSE}. 
Being this latest work the most applicable describing the
dynamics of BECs. 
Some examples of 1D and 2D reduction for studying BECs
with a non-local dipolar interaction 
in a mean-field approximation are given in 
\cite{Dim-red-dip}.
A starting point for the theoretical study of SO coupled 
BECs in reduced dimensionality is provided by a binary
mean-field nonpolynomial Schr\"odinger equation (NPSE), 
which is derived in \cite{Bright-4}.
The system of coupled equations was used to investigate
localized modes in dense repulsive and attractive BECs 
with the SO and Rabi couplings. Recently, it was considered
an effectively 2D BEC with the SO coupling of mixed
Rashba-Dresselhaus type and Rabi term \cite{Salasnich-2D}.
The system is described by two coupled 2D NPSEs,
for both attractive and repulsive inter-atomic interactions.
Approximate localized solutions are analyzed 
by treating the SO and Rabi terms as perturbations. 
Localized solutions of the system are obtained in a 
numerical form. 
There have also been theoretical efforts toward 
understanding the physics of the SO coupled Bose gases 
at finite temperature \cite{T-finite}. 
Here it was found that in three spatial dimensions SO 
coupling lowers the critical temperature of condensation
and enhances thermal depletion of the condensate fraction. 
In two dimensions the SO coupling destroys superfluidity
at any finite temperature.

In this paper, starting from the 3D many-body Hamiltonian
describing $N$ interacting bosons with SO and Rabi couplings
and where the dynamics is ruled by two coupled field 
equations, we investigate the 1D and 2D reduction of a
bosonic quantum field theory.
In the derivation of the 1D model we use a transverse
and isotropic harmonic confinement and a generic trapping 
potential in the axial direction.
In the 2D system the trapping potential is harmonic in the
axial direction and generic in the transverse one.
We achieve to getting reduced models of 
time-dependent 1D and 2D nonpolynomial Heisenberg coupled 
equations. 
The derivation is performed for both the repulsive and 
attractive signs of the inter-atomic, intra- and 
inter-species interactions. 
In the case which the many-body quantum
state of the system coincides with the Glauber 
coherent state our findings agree with the two 
coupled NPSEs in 1D \cite{Bright-4} and
2D \cite{Salasnich-2D}.
The derivation of the effective equations is followed 
by the numerical analysis of new features of trapped modes
in the 1D NPSE and 1D GPE describing the dynamics 
of a BEC with SO and Rabi couplings, 
for both signs of the nonlinearity.

The rest of the paper is organized in the following way.
In the Sec. \ref{II}, we derive an effective 1D 
nonpolynomial Heisenberg equation. 
This section also includes the condition under which the
Heisenberg equation coincides with the 1D NPSE
describing a BEC with SO and Rabi couplings.
The 2D reduction of the 3D  Hamiltonian is presented
in Sec. \ref{III}.
Numerical results of the 1D NPSE and 1D GPE
are reported in Sec. \ref{IV}.
Finally, we present a summary and discussion 
of our study in Sec. \ref{V}.

\section{Dimensional reduction of a quantum field theory}
\label{II}

A quantum treatment of interacting bosons with SO and Rabi
couplings can be obtained starting from the many-body
Hamiltonian in second quantization, which describes
$N$ interacting bosons 
of equal mass $m$ confined by an external potential
$U\left(\mathbf{r}\right)$,
\begin{eqnarray} 
{\hat H} &=& \int d{\bf r} \ 
{\hat \Psi}^{\dagger} \left(\mathbf{r}\right)
H_{sp} {\hat \Psi}\left(\mathbf{r}\right)
\nonumber \\
&+& \frac{1}{2} \int d{\bf r} \
{\hat \Psi}^{\dagger} \left(\mathbf{r}\right)
{\hat \Psi}^\dagger \left(\mathbf{r'}\right)
V\left(\mathbf{r-r'}\right)
{\hat \Psi} \left(\mathbf{r}'\right)
{\hat \Psi} \left(\mathbf{r}\right)
\label{Hamiltionian}
\end{eqnarray} 
where ${\hat \Psi}^\dagger \left(\mathbf{r}\right)$ and
${\hat \Psi}\left(\mathbf{r}\right)$ are the 
two pseudo-spin creation and annihilation boson field
operators, respectively. Thus, 
${\hat \Psi}  =
({\hat \psi_{1}} ,\,\, {\hat \psi_{2}} )^{T}$
and ${\hat \Psi}^\dagger =
({\hat \psi_{1}}^\dagger \,\,\, {\hat \psi_{2}}^\dagger)$.
The term $V\left(\mathbf{r-r'}\right)$ stands for
the two-body interaction.
The single-particle SO Hamiltonian $H_{sp}$ \cite{Bright-4}
is,
\begin{eqnarray}
H_{sp}= \left[ \frac{\mathbf{\hat{p}}^{2}}{2m} + U\left(\mathbf{r}\right)
\right]\sigma_{0} + \frac{\hbar \Omega}{2}\sigma_{x} - 
\frac{k_{L}}{m} \ \hat{p}_{x} \sigma_{z}
\label{H_original}
\end{eqnarray}
where $\mathbf{\hat {p}}^{2}=-\hbar^{2} \nabla^{2}$ is the
square of the momentum operator, $U\left(\mathbf{r}\right)$
is a trapping potential, $\Omega$ is the frequency of the
Raman coupling, which is responsible for the Rabi mixing
between the two states, $k_{L}$ is the recoil wave 
number induced by 
the interaction with the laser beams, $\sigma_{0}$ is the 
two-dimensional unit matrix and $\sigma_{x,z}$ are the Pauli
matrices.
By assuming a dilute system of bosons, the inter-particle
interaction can be approximate  by a contact
pseudo-potential. 
We consider interactions for both intra- and 
inter-species. This means that the potential
for intra-species interactions is assumed as
$V({\bf r- r}') 
\equiv (4\pi \hbar^2 a_{jj}/m )\delta({\bf r- r}')$,
where $4\pi \hbar^2 a_{jj}/m$ 
is the strength of interaction, $a_{jj}$ is the 
intra-species s-wave scattering length with 
$j=1,2$ for the two pseudo-spin components. 
In the inter-species interactions the potential
has the same form, but the
strength of the interaction takes the form 
$4\pi \hbar^2 a_{12}/m$. 
In order to transform the Hamiltonian (\ref{Hamiltionian})
into dimensionless form, we use the HO length of the 
transverse trap $a_{\perp}=\sqrt{\hbar/(m\omega_{\perp})}$,
with the trapping frequency $\omega_{\perp}$.
The time $t$ is measured in units of $\omega^{-1}_{\perp}$, 
the spatial variable ${\bf r}$ in units of $a_{\perp}$,
the energy in units $\hbar \omega_{\perp}$ and the field 
operators are given in units of $a_{\perp}^{3/2}$.
By using these new variables we have the dimensionless
many-body Hamiltonian in second quantization,
\begin{eqnarray}
{\hat H} = {\hat H}_{sp} + {\hat H}_{int}
\label{H-field}
\end{eqnarray}
where,
\begin{eqnarray} 
{\hat H}_{sp} &=& \int d{\bf r}
\Bigg \{\sum_{j=1,2} {\hat \psi}_{j}^\dagger
\Bigg [ -{\frac{1}{2}}\nabla^{2}
+ U(\mathbf{r}) 
\nonumber \\
&+& (-1)^{j-1}{\mbox i} \gamma \frac{\partial}{\partial x}
\Bigg] {\hat \psi}_{j}
+ \Gamma \big ({\hat \psi}_{1}^\dagger {\hat \psi}_{2} +
{\hat \psi}_{2}^\dagger {\hat \psi}_{1} \big) \Bigg \}
\label{H_sp}
\end{eqnarray}
and 
\begin{eqnarray} 
{\hat H}_{int} &=& \int d{\bf r}
\Bigg [\sum_{j=1,2} \pi g_{jj} {\hat \psi}_{j}^\dagger 
{\hat \psi}_{j}^\dagger
{\hat \psi}_{j} {\hat \psi}_{j} 
+ 2\pi g_{12} {\hat \psi}_{1}^\dagger {\hat \psi}_{2}^\dagger
{\hat \psi}_{2} {\hat \psi}_{1} \Bigg]
\label{H_int}
\end{eqnarray}
where $j=(1,2)$ and $\psi_{j}$ 
represents the field operators of the two atomic states,
$g_{jj}\equiv 2a_{jj}/a_{\perp}$, 
$g_{12}\equiv 2a_{12}/a_{\perp}$ are the strengths of
the intra- and inter-species interactions. 
$\gamma \equiv k_{L}a_{\perp}$ and 
$\Gamma  \equiv  \Omega /(2\omega_{\perp})$ are dimensionless
strengths of the SO and Rabi couplings, respectively. 
The external potential is given as
$U(\mathbf{r})= V(x) + (y^2 + z^2)/2$.
Thus, the many-body Hamiltonian (\ref{H-field}) describes a
dilute gas of bosonic atoms with SO and Rabi couplings 
confined in the transverse $(y, z)$ plane by a HO
potential and a generic potential $V(x)$ in the $x$ axial
direction.
The bosonic field operator 
${\hat \psi}({\bf r},t)$ and its adjoint
${\hat \psi}^{\dagger}({\bf r},t)$
must satisfy the following equal-time commutation rules 
\begin{equation}
[ {\hat \psi}_{\alpha}({\bf r},t) , 
{\hat \psi}_{\beta}^{\dagger}({\bf r}',t) ] =
\delta({\bf r}-{\bf r}')\delta_{\alpha \beta} 
\label{commuta-Bose-1}
\end{equation}
\begin{equation}
[ {\hat \psi}_{\alpha}({\bf r},t) ,
{\hat \psi}_{\beta}({\bf r}',t) ] = 
[ {\hat \psi}_{\alpha}^{\dagger}({\bf r},t) , 
{\hat \psi}_{\beta}^{\dagger}({\bf r}',t) ] = 0
\label{commuta-Bose-2}
\end{equation}
where $\alpha, \beta =1,2$.
By imposing these commutation rules we find  the creation of a
particle in the state $|{\bf r},\alpha,t\rangle$ from the vacuum
state $|0\rangle$ as, 
${\hat \psi}_{\alpha}^{\dagger}({\bf r},t) |0\rangle = |{\bf r},\alpha,t\rangle$.
The annihilation of a particle, which is in the state 
$|{\bf r},\beta,t\rangle$ is given by
${\hat \psi}_{\alpha}({\bf r}',t) |{\bf r},\beta, t\rangle = 
\delta({\bf r}-{\bf r}')\delta_{\alpha\beta} |0\rangle$. 
From the Heisenberg equation of motion 
\begin{eqnarray}
{\mbox i} {\partial {\hat \Psi}  \over \partial t} 
= [ {\hat \Psi} , {\hat H} ] \;  
\label{Heisenberg-eq}
\end{eqnarray}
we get two coupled field equations in a closed form,
\begin{eqnarray}
{\mbox i}
{\frac{\partial}{\partial t}}{\hat \psi}_{j}(\mathbf{r},t) 
&=&
\Bigg[ -{\frac{1}{2}}\nabla^{2} + V(x)
+ {1\over 2} \left( y^2 + z^2 \right)
\nonumber 
\\
&+& (-1)^{j-1}i \gamma \frac{\partial}{\partial x}
+2\pi g_{jj} {\hat \psi}_{j}^{\dagger}(\mathbf{r},t)
{\hat \psi}_{j}(\mathbf{r},t)
\nonumber 
\\
&+& 2\pi g_{12} {\hat \psi}_{3-j}^{\dagger}(\mathbf{r},t)
{\hat \psi}_{3-j}(\mathbf{r},t)\Bigg] {\hat \psi}_{j}
(\mathbf{r},t) 
+\Gamma {\hat \psi}_{3-j}(\mathbf{r},t)
\label{field-eq}
\end{eqnarray}

\subsection{1D-reduction of the 3D-Hamiltonian}

To perform the one-dimensional reduction of the 3D 
Hamiltonian (\ref{H-field}) we suppose that in the
transverse $(y, z)$ plane the single-particle ground-state
is a Gaussian wave-function 
\cite{Salasnich-Quantum-solitons, Salasnich-QFT,
Salasnich-book} as in a BEC. Thus
\begin{equation}
{\hat \psi}_{j}({\bf r}) |G\rangle = 
{1\over \sqrt\pi \ \eta_{j}(x,t)}
\exp{\left[ - {y^2+z^2\over 2\eta^2_{j}(x,t)} 
\right] }\, {\hat \phi}_{j}(x,t) |G\rangle \; 
\label{supposing-1}
\end{equation}
where $|G\rangle$ is the many-body ground state, while 
$\eta_{j}(x,t)$ is the transverse width for
bosonic field operators and
${\hat \phi}_{j}(x,t)= ({\hat \phi}_{1}(x,t), 
{\hat \phi}_{2}(x,t))^T$
represents the axial bosonic field operators.
By applying this ansatz into the Hamiltonian
(\ref{H-field}), we obtain the 
effective Hamiltonian ${\hat h}_{1D}$ such that
\begin{eqnarray}
{\hat H}|G\rangle = {\hat h}_{1D} |G\rangle 
\end{eqnarray}
Neglecting the space derivatives of $\eta_{j}(x,t)$,
the effective 1D-Hamiltonian can be read
\begin{eqnarray} 
{\hat h}_{1D} &=& \int dx \
\Bigg \{\sum_{j=1,2} {\hat \phi}_{j}^\dagger
\Bigg [ -{\frac{1}{2}} \frac{\partial^2}{\partial x^2}
+ V(x) + \frac{1}{2} \left( \frac{1}{\eta_{j}^2} + 
\eta_{j}^2 \right)
\nonumber \\
&+& (-1)^{j-1}{\mbox i} \gamma \frac{\partial}{\partial x} 
+\frac{g_{jj}}{2\eta^{2}_{j}}{\hat \phi}_{j}^\dagger 
{\hat \phi}_{j}\Bigg]
{\hat \phi}_{j} + \frac{2 g_{12}}{\eta^2_{1} + \eta^2_{2}}
{\hat \phi}_{1}^\dagger {\hat \phi}_{2}^\dagger
{\hat \phi}_{2} {\hat \phi}_{1}
\nonumber \\
&+& 2\Gamma \frac{\eta_{1} \eta_{2}}{\eta_{1}^2 +
\eta_{2}^2} \sum_{j=1,2} 
{\hat \phi}_{j}^\dagger {\hat \phi}_{3-j}
\Bigg\}
\label{h}
\end{eqnarray}
The ground state $|G\rangle$ is obtained self-consistently 
from the Hamiltonian (\ref{h}) and the transverse width 
$\eta_{j}(x,t)$. It is determined by minimizing the
energy functional $\langle G|{\hat h}|G\rangle$ with
respect to $\eta_{j}(x,t)$. In this way we get
\begin{eqnarray}
\eta_{j}^4 &=&
1+ g_{jj} \frac{\langle G| {\hat \phi}_{j}^\dagger
{\hat \phi}_{j}^\dagger
{\hat \phi}_{j} {\hat \phi}_{j} |G\rangle } {\langle G|
{\hat \phi}_{j}^\dagger {\hat \phi}_{j} |G\rangle}
+ 4g_{12} \frac{\eta_{j}^{4}}{(\eta^2_{1} + \eta^2_{2})^2}
\frac{\langle G| {\hat \phi}_{1}^\dagger {\hat \phi}_{2}^\dagger
{\hat \phi}_{2} {\hat \phi}_{1} |G\rangle } {\langle G|
{\hat \phi}_{j}^\dagger {\hat \phi}_{j} |G\rangle}
\nonumber \\
&+& 2\Gamma (-1)^{j-1} \eta_{j}^{3} \eta_{3-j} 
\frac{\eta_{1}^2 - \eta_{2}^2}{(\eta_{1}^2 + \eta_{2}^2)^2}
\frac {\langle G| ({\hat \phi}_{1}^\dagger {\hat \phi}_{2} + 
{\hat \phi}_{2}^\dagger {\hat \phi}_{1}) |G\rangle}
{\langle G|{\hat \phi}_{j}^\dagger {\hat \phi}_{j} |G\rangle}
\end{eqnarray}
The 1D-Hamiltonian ${\hat h}_{1D}$ (\ref{h}), is a 
generalization of the one introduced by Salasnich 
\cite{Salasnich-QFT}, in the context of a BEC in
presence of an axial potential.
In the case which the SO and Rabi couplings are dropped the
effective Hamiltonian ${\hat h}_{1D}$ corresponds with 
the model implemented in \cite{Salasnich-QFT}.

\subsection{1D-nonpolynomial Heisenberg equation}

From the effective 1D-Hamiltonian (\ref{h}) and the
Heisenberg equation of motion 
\begin{equation} 
{\mbox i} {\partial \over \partial t} {\hat \phi}_{j} 
= [ {\hat \phi}_{j} , {\hat h}_{1D} ] \;  
\end{equation} 
we derive the 1D-nonpolynomial Heisenberg equation,
which represents the motion of a many-body quantum
system of dilute bosonic atoms,
\begin{eqnarray} 
{\mbox i} {\partial \over \partial t} {\hat \phi}_{j} &=&
\Bigg [ -{\frac{1}{2}} \frac{\partial^2}{\partial x^2}
+ V(x) + \frac{1}{2} \left( \frac{1}{\eta_{j}^2} + 
\eta_{j}^2 \right)
+ (-1)^{j-1}{\mbox i} \gamma \frac{\partial}{\partial x} 
+ \frac{g_{jj}}{\eta^{2}_{j}} 
{\hat \phi}_{j}^\dagger {\hat \phi}_{j} 
\nonumber \\
&+&\frac{2 g_{12}}{\eta^2_{1} + \eta^2_{2}}
{\hat \phi}_{3-j}^\dagger {\hat \phi}_{3-j} \Bigg]
{\hat \phi}_{j}
+ 2\Gamma \frac{\eta_{1} \eta_{2}}{\eta_{1}^2 + \eta_{2}^2}
{\hat \phi}_{3-j}
\label{1DNPSE-field}
\end{eqnarray}
This equation must be solved self-consistently by using the
many-body quantum state of the system $|S\rangle$ in the
equations of the transverse widths $\eta_{j}$. So
\begin{eqnarray}
\eta_{j}^4 (x,t) &=&
1+ g_{jj} \frac{\langle S| {\hat \phi}_{j}^\dagger (x,t)
{\hat \phi}_{j}^\dagger(x,t)
{\hat \phi}_{j} (x,t){\hat \phi}_{j}(x,t) |S\rangle } 
{\langle S|
{\hat \phi}_{j}^\dagger(x,t) {\hat \phi}_{j}(x,t) |S\rangle}
\nonumber \\
&+& 4g_{12} \frac{\eta_{j}^{4}}{(\eta^2_{1} + \eta^2_{2})^2}
\frac{\langle S| {\hat \phi}_{1}^\dagger(x,t) {\hat \phi}_{2}^\dagger(x,t)
{\hat \phi}_{2} (x,t){\hat \phi}_{1} (x,t)|S\rangle } 
{\langle S|
{\hat \phi}_{j}^\dagger (x,t){\hat \phi}_{j} (x,t)|S\rangle}
\nonumber \\
&+& 2\Gamma (-1)^{j-1} \eta_{j}^{3} \eta_{3-j} 
\frac{\eta_{1}^2 - \eta_{2}^2}{(\eta_{1}^2 + \eta_{2}^2)^2}
\nonumber \\
&\times&
\frac {\langle S| ({\hat \phi}_{1}^\dagger(x,t) 
{\hat \phi}_{2}(x,t) + {\hat \phi}_{2}^\dagger(x,t) 
{\hat \phi}_{1}(x,t)) |S\rangle}
{\langle S|{\hat \phi}_{j}^\dagger(x,t) {\hat \phi}_{j}(x,t)
|S\rangle}
\end{eqnarray}
In an open environment particles may be added or removed
from the condensate, and one can suppose that the number 
of bosons in the matter field is not fixed. This 
implies that the condensate is not in a pure Fock state
\cite{Salasnich-book,Coherent-states-QFT,
Coherent-states-1}.
We then introduce, in analogy with optics the Glauber
coherent states $|GCS\rangle$ \cite{Coherent-states-2},
which do not have a fixed number of particles.
On the other hand, the interacting many-body system
described with quantum field theory is equivalent to an
infinite number of interacting harmonic oscillators.
As an useful ingredient in the functional formulation of 
quantum field theory \cite{Coherent-states-QFT},
it is possible introduce the coherent
states in the familiar setting of a single harmonic 
oscillator.
In this theory, the partition function is writing as a 
functional integral over time-dependent fields, 
which are the eigenvalues of the coherent states.
The Glauber coherent state $|GCS\rangle$, is introduced 
as the eigenstate of the annihilation operator.
In terms of quantum field operators we have
${\hat \phi}_{j}(x,t)|GCS\rangle = \phi_{j}(x,t)|GCS\rangle$
where $j=1,2$ and $\phi_{j}(x,t)$ is a classical field.
The average number of atoms in the coherent state is
given by $N_{j}=\langle GCS|\hat N_{j}|GCS\rangle$.
Hence in the case in which the many-body state
$|S\rangle$ coincides with the Glauber coherent state 
$|GCS\rangle$ 
\cite{Salasnich-Quantum-solitons, Salasnich-QFT},
the 1D-nonpolynomial Heisenberg equation 
(\ref{1DNPSE-field}) becomes 
1D-nonpolynomial Schr\"odinger equation describing a
BEC with SO and Rabi couplings
\begin{eqnarray} 
{\mbox i} {\partial \over \partial t} \phi_{j} &=&
\Bigg [ -{\frac{1}{2}} \frac{\partial^2}{\partial x^2}
+ V(x) + \frac{1}{2} \left( \frac{1}{\eta_{j}^2} + 
\eta_{j}^2 \right)
+ (-1)^{j-1}{\mbox i} \gamma \frac{\partial}{\partial x}
\nonumber \\
&+&
\frac{g_{jj}} {\eta^{2}_{j}} \left|\phi_{j}\right|^2
+\frac{2 g_{12}}{\eta^2_{1} + \eta^2_{2}}
\left|\phi_{3-j}\right|^2 \Bigg]
\phi_{j}
+ 2\Gamma \frac{\eta_{1} \eta_{2}}{\eta_{1}^2 +
\eta_{2}^2} \phi_{3-j}
\label{NPSE-SOC}
\end{eqnarray}
where $\phi_{j}(x,t)$ is a complex wave-function, with the 
normalization condition 
$\int_{-\infty}^{\infty}dx|\phi_{j}(x,t)|^2=N_{j}$, and
the conserved total number of atoms is $N=N_{1}+N_{2}$.
The corresponding transverse widths $\eta_{j}(x,t)$, are
\begin{eqnarray}
\eta_{j}^4 &=&
1+ g_{jj} \left|\phi_{j}\right|^2
+ 4g_{12} \frac{\eta_{j}^{4}}{(\eta^2_{1} + \eta^2_{2})^2}
\left|\phi_{3-j}\right|^2
\nonumber \\
&+& 2\Gamma (-1)^{j-1} \eta_{j}^{3} \eta_{3-j} 
\frac{\eta_{1}^2 - \eta_{2}^2}{(\eta_{1}^2 + \eta_{2}^2)^2}
\Bigg(\frac { \phi_{1}^{*} \phi_{2} + \phi_{2}^{*} \phi_{1}}
{\left|\phi_{j}\right|^2 }\Bigg)
\label{widths}
\end{eqnarray}
The models (\ref{NPSE-SOC}) and (\ref{widths}) were
derived previously by means of a mean-field approximation 
\cite{Bright-4}. 
Here the Bogoliubov approximation \cite{Dalfovo} is applied
in the Hamiltionian (\ref{H-field}), according to
which since the condensate state involves the macroscopic
occupation of a single state it is appropriate to split the
Bose field operator in two terms, 
the first one describes a macroscopically
populated state, defined as the expectation value
of the field operator, 
and the second one takes into account quantum
fluctuations about the state condensed.
Since the number of atoms in the single-particle condensate 
wavefunction is large, the quantum fluctuations are
negligible and the dimensional reduction
\cite{Salasnich-NPSE} of resulting Hamiltonian 
leads to the mean-field approximation given above.
Our results let see the correspondence between
quantum field theory and classical field theory.
Besides, these suggest that the models obtained by means 
of Glauber coherent state are
indeed an equivalent tool to the Bogoliubov approximation
in research of BECs at zero temperature.
In the symmetric case, when the effective nonlinearity
coefficients for the intra- and inter-species interactions
are equal, i.e., $g_{11}=g_{22}=g_{12}\equiv g$, 
we have $\eta_{1}=\eta_{2}$ and Eq. (\ref{widths}) 
takes the form
$\eta_{1}^4=\eta_{2}^4= 1+ g(\left|\phi_{1}\right|^2 + 
\left|\phi_{2}\right|^2)$. 
We note that, for the self-repulsive binary BEC $g>0$,
and the self-attractive binary BEC $g<0$.
In that way, the system of two coupled equations
(\ref{NPSE-SOC}) represents a generalization of the model 
introduced earlier for the study of vectorial solitons in
two-component BECs \cite{vector-solit} and it can be
read as \cite{Bright-4}
\begin{eqnarray} 
{\mbox i} {\partial \over \partial t} \phi_{j} &=&
\Bigg [ -{\frac{1}{2}} \frac{\partial^2}{\partial x^2}
+ V(x) + (-1)^{j-1}{\mbox i}
\gamma \frac{\partial}{\partial x} 
\nonumber \\
&+& \frac{1 + (3/2)g(\left|\phi_{1}\right|^2 + 
\left|\phi_{2}\right|^2)}{\sqrt{1 + g(\left|\phi_{1}\right|^2 + 
\left|\phi_{2}\right|^2)}} \Bigg] \phi_{j} + \Gamma 
\phi_{3-j}
\label{NPSE-SO-Rabi}
\end{eqnarray}
Only if $g(|\phi_{1}|^2 + |\phi_{2}|^2)\ll 1$  
may the system (\ref{NPSE-SO-Rabi}) be considered as 
one-dimensional.
It is possible to construct stationary states using Eq. 
(\ref{NPSE-SO-Rabi}). These states are constructed with the 
chemical potential $\mu$, by setting
$\phi_{j}(x,t)\rightarrow \phi_{j}(x)\exp{(-i\mu t)}$.
The resulting equations for stationary fields $\phi_{1,2}$
are compatible with the restriction 
$\phi_{1}^*(x)=\phi_{2}(x)$.  In terms of real 
$\mathcal{R}$ and imaginary $\mathcal{I}$ part we have 
$\mathcal{R}\{\phi_{1}\}=\mathcal{R}\{\phi_{2}\}$, and
$\mathcal{I}\{\phi_{1}\}=-\mathcal{I}\{\phi_{2}\}$.
Thus leading to a single stationary NPSE \cite{Bright-4},
\begin{eqnarray} 
\mu \Phi &=&
\Bigg [ -{\frac{1}{2}} \frac{d^2}{d x^2}
+ V(x) + i \gamma \frac{d}{d x} 
+\frac{1 + (3/2)gN |\Phi|^2}{\sqrt{1 + gN |\Phi|^2}}
\Bigg] \Phi + \Gamma \Phi^{*}
\label{Stationary-NPSE}
\end{eqnarray}
where we have defined $\Phi(x)\equiv \sqrt{2/N}\phi_{1}(x)$,
along with the condition mentioned above between 
the stationary fields $\phi_{1,2}$. The normalization
becomes $\int_{-\infty}^{\infty} dx |\Phi(x)|^2=1$.
In the strong coupling regime $gN|\Phi|^2 \gg 1$, the
nonpolynomial nonlinearity reduces to a quadratic form
$(3/2)\sqrt{gN}|\Phi|\Phi$. The other hand, in the weakly
nonlinear regime $gN|\Phi|^2 \ll 1$ the
nonlinearity takes the cubic form $\sim gN|\Phi|^2\Phi$. 
Where without loss the generality, it is omitted 
the transverse contribution.
In general, the coupled equations (\ref{NPSE-SOC}) 
are strictly one-dimensional \cite{GPE-SOC-1D},
under the condition $\eta_{j}^4= 1$ in the system of Eqs.
(\ref{widths}), establishing the 1D GPE with SO and Rabi 
couplings
\begin{eqnarray} 
{\mbox i} {\partial \over \partial t} \phi_{j} &=&
\Bigg [ -{\frac{1}{2}} \frac{\partial^2}{\partial x^2}
+ V(x) + (-1)^{j-1}{\mbox i}\gamma \frac{\partial}{\partial x} 
\nonumber \\
&+& g_{jj} \left|\phi_{j}\right|^2
+ g_{12} \left|\phi_{3-j}\right|^2 \Bigg] \phi_{j} 
+\Gamma \phi_{3-j}
\label{1D-eq}
\end{eqnarray}
we have omitted the constant contribution of the transverse 
energy given by $1$ (in units of $\omega_{\perp}$).
By performing a global pseudo-spin
rotation $\sigma_{x}\rightarrow \sigma_{z}$
and $\sigma_{z}\rightarrow \sigma_{x}$, in the 
Hamiltonian (\ref{H_original}), the model (\ref{1D-eq})
becomes the one that was implemented to studying of nonlinear
modes in binary bosonic condensates with nonlinear
repulsive interactions and linear SO- and 
Zeeman-splitting couplings \cite{Malomed-1}.

\section{2D-reduction of the 3D-Hamiltonian}
\label{III}

In analogy with the 1D scenario, we use the 
3D-Hamiltonian (\ref{Hamiltionian}) to obtain an 
effective 2D-Hamiltonian ${\hat h}_{2D}$ and the respective
2D-nonpolynomial Heisenberg equation. 
The 3D-Hamiltonian is scaled by means of HO length
$a_{z}=\sqrt{\hbar/(m\omega_{z})}$, which is
established by the trap in the $z$ axis.
The length, time, energy and field operators
are given in units of
$a_{z}$, $\omega_{z}^{-1}$, $\hbar \omega_{z}$,
and $a_{z}^{3/2}$, respectively. Thus, 
$ {\hat H}$ is given by
\begin{eqnarray} 
{\hat H} &=& \int d{\bf r}
\Bigg \{\sum_{j=1,2} {\hat \psi}_{j}^\dagger
\Bigg [ -{\frac{1}{2}}\nabla^{2}
+ V(x,y) + \frac{1}{2}z^2
+(-1)^{j-1}{\mbox i}  \gamma \frac{\partial}{\partial x}
\nonumber \\
&+& \pi g_{jj} {\hat \psi}_{j}^\dagger{\hat \psi}_{j}\Bigg]
{\hat \psi}_{j}
+ 2\pi g_{12} {\hat \psi}_{1}^\dagger {\hat \psi}_{2}
^\dagger
{\hat \psi}_{2} {\hat \psi}_{1}
+ \Gamma \sum_{j=1,2} {\hat \psi}_{j}^\dagger
{\hat \psi}_{3-j}\Bigg \}
\label{H-field-2}
\end{eqnarray}
where $j=(1,2)$ and $\psi_{j}$ 
represents the field operators of the two atomic states,
$g_{jj}\equiv 2a_{jj}/a_{z}$, 
$g_{12}\equiv 2a_{12}/a_{z}$ are the strengths of the intra- 
and inter-species interactions. 
While, $\gamma \equiv k_{L}a_{z}$ and 
$\Gamma  \equiv  \Omega /(2\omega_{z})$ are dimensionless
strengths of the SO and Rabi couplings, respectively. 
The external potential is given by $V(x,y) + z^2/2$.
Thus, the many-body Hamiltonian (\ref{H-field-2}) describes
a dilute gas of bosonic atoms with SO and Rabi couplings 
confined in the axial $z$ direction by a HO
potential and a generic potential $V(x,y)$ in the transverse
$(x,y)$ plane. The bosonic field operators satisfy the
commutation relations (\ref{commuta-Bose-1}) and 
(\ref{commuta-Bose-2}).
The effective 2D-Hamiltonian ${\hat h}_{2D}$ is derived by 
supposing a Gaussian wave-function for the single-particle
ground-state in the $z$ axis, as in a BEC. Similarly,
as it was done above in the 1D case,
\begin{eqnarray}
{\hat \psi}_{j}({\bf r}) |G\rangle &=& 
{1\over \bigg [\pi \xi_{j}^2(x,y,t)\bigg ]^{1/4}}
\exp{\left[ - {z^2\over 2\ \xi^2_{j}(x,y,t)} 
\right] }\, {\hat \phi}_{j}(x,y,t) |G\rangle \; 
\label{supposing-1}
\end{eqnarray}
where $|G\rangle$ is the many-body ground state, while 
$\xi_{j}(x,y,t)$ is the axial width for
bosonic field operators and
${\hat \phi}_{j}(x,y,t)= ({\hat \phi}_{1}(x,y,t), 
{\hat \phi}_{2}(x,y,t))^T$
represents the transverse bosonic field operators.
By applying this ansatz into the Hamiltonian 
(\ref{H-field-2}) and neglecting the space derivatives of 
$\xi_{j}(x,y,t)$, we obtain the effective Hamiltonian
${\hat h}_{2D}$ as
${\hat H}|G\rangle = {\hat h}_{2D} |G\rangle$, with
\begin{eqnarray} 
{\hat h}_{2D} &=& \int dx dy
\Bigg \{\sum_{j=1,2} {\hat \phi}_{j}^\dagger
\Bigg [ -{\frac{1}{2}} \Bigg (\frac{\partial^2}
{\partial x^2}+\frac{\partial^2}{\partial y^2} \Bigg)+V(x,y) 
\nonumber \\
&+& \frac{1}{4} \left( \frac{1}{\xi_{j}^2}+\xi_{j}^2\right)
+ (-1)^{j-1}{\mbox i} \gamma \frac{\partial}{\partial x}
+ \sqrt{\frac{\pi}{2}}\frac{g_{jj}}{\xi_{j}} 
{\hat \phi}_{j}^\dagger {\hat \phi}_{j}\Bigg] {\hat \phi}_{j}
\nonumber \\
&+& \frac{2 \sqrt{\pi} g_{12}}{\sqrt{\xi^2_{1}+\xi^2_{2}}}
{\hat \phi}_{1}^\dagger {\hat \phi}_{2}^\dagger
{\hat \phi}_{2} {\hat \phi}_{1} 
+ \Gamma \sqrt{\frac{2\xi_{1} \xi_{2}}{\xi_{1}^2 + 
\xi_{2}^2}}\sum_{j=1,2} {\hat \phi}_{j}^\dagger 
{\hat \phi}_{3-j}\Bigg\}
\label{h_2D}
\end{eqnarray}
and by minimizing the energy functional 
$\langle G|{\hat h}_{2D}|G\rangle$ with respect to
$\xi_{j}(x,y,t)$, we determine the axial widths 
$\xi_{j}(x,y,t)$.  
The ground state $|G\rangle$ is obtained self-consistently 
from the Hamiltonian (\ref{h_2D}) and the axial widths
$\xi_{j}$, 
\begin{eqnarray}
\xi_{j}^4 &=&
1+ \sqrt{2\pi}g_{jj}\xi_{j}\frac{\langle G| 
{\hat \phi}_{j}^\dagger {\hat \phi}_{j}^\dagger
{\hat \phi}_{j}
{\hat \phi}_{j} |G\rangle } {\langle G|
{\hat \phi}_{j}^\dagger {\hat \phi}_{j} |G\rangle}
+ 4\sqrt{\pi}g_{12} \frac{\xi_{j}^{4}}{(\xi^2_{1} +
\xi^2_{2})^{3/2}}
\frac{\langle G| {\hat \phi}_{1}^\dagger 
{\hat \phi}_{2}^\dagger
{\hat \phi}_{2} {\hat \phi}_{1} |G\rangle } {\langle G|
{\hat \phi}_{j}^\dagger {\hat \phi}_{j} |G\rangle}
\nonumber \\
&+& \sqrt{2}\Gamma (-1)^{j-1} \xi_{j}^{5/2} \xi^{1/2}_{3-j}
\frac{\xi_{1}^2 - \xi_{2}^2}{(\xi_{1}^2 + \xi_{2}^2)^{3/2}}
\frac {\langle G| ({\hat \phi}_{1}^\dagger {\hat \phi}_{2} + 
{\hat \phi}_{2}^\dagger {\hat \phi}_{1}) |G\rangle}
{\langle G|{\hat \phi}_{j}^\dagger {\hat \phi}_{j} |G\rangle}
\label{width-2D}
\end{eqnarray}

\subsection{2D-nonpolynomial Heisenberg equation}

By means of the 2D-Hamiltonian ${\hat h}_{2D}$ and the 
Heisenberg equation
${\mbox i} 
{\partial_{t}} {\hat \phi}_{j} = [ {\hat \phi}_{j} , 
{\hat h}_{2D}]$, we derive the 2D-nonpolynomial Heisenberg 
equation, 
\begin{eqnarray} 
{\mbox i}  {\partial \over \partial t} {\hat \phi}_{j} &=&
\Bigg [ -{\frac{1}{2}} \bigg (\frac{\partial^2}
{\partial x^2} + \frac{\partial^2}{\partial y^2} \bigg)
+ V(x,y) + \frac{1}{4} \left( \frac{1}{\xi_{j}^2} 
+ \xi_{j}^2 \right)
+ (-1)^{j-1}{\mbox i}  \gamma \frac{\partial}{\partial x} 
\nonumber \\
&+& \sqrt{2\pi}\frac{g_{jj}}{\xi_{j}} 
{\hat \phi}_{j}^\dagger {\hat \phi}_{j} 
+ \frac{2\sqrt{\pi}g_{12}}{\sqrt{\xi^2_{1} + \xi^2_{2}}}
{\hat \phi}_{3-j}^\dagger {\hat \phi}_{3-j} \Bigg]
{\hat \phi}_{j}
+ \Gamma \sqrt{\frac{2\xi_{1} \xi_{2}}{\xi_{1}^2+\xi_{2}^2}}
{\hat \phi}_{3-j}
\label{2DNPSE-field}
\end{eqnarray}
which must be solved self-consistently by using the
many-body quantum state of the system $|S\rangle$ in the
equations of the transverse widths $\xi_{j}$.  
So in the system  (\ref{width-2D}) we need changing
the ground state by the  many-body quantum state, i.e. 
$|G\rangle \rightarrow |S\rangle$.
When the many-body state $|S\rangle$ 
coincides with the Glauber coherent state $|GCS\rangle$ 
\cite{Salasnich-Quantum-solitons, Salasnich-QFT},
such that, 
${\hat \phi}_{j}(x,y,t)|GCS\rangle 
= \phi_{j}(x,y,t)|GCS\rangle$,
the 2D-nonpolynomial Heisenberg equation becomes 2D-NPSE
describing a BEC with SO and Rabi couplings in a 
mean-field approximation
\begin{eqnarray} 
{\mbox i}  {\partial \over \partial t} \phi_{j} &=&
\Bigg [ -{\frac{1}{2}} \bigg (\frac{\partial^2}
{\partial x^2}+\frac{\partial^2}{\partial y^2} \bigg)
+ V(x,y) + \frac{1}{4} \left( \frac{1}{\xi_{j}^2} +
\xi_{j}^2 \right)
+ (-1)^{j-1}{\mbox i}  \gamma \frac{\partial}{\partial x} 
\nonumber \\
&+& \sqrt{2\pi}\frac{g_{jj}}{\xi_{j}} 
\left| \phi_{j} \right|^2
+\frac{2\sqrt{\pi}g_{12}}{\sqrt{\xi^2_{1} + \xi^2_{2}}}
|\phi_{3-j}|^2 \Bigg] \phi_{j}
+ \Gamma \sqrt{\frac{2\xi_{1} \xi_{2}}{\xi_{1}^2 + 
\xi_{2}^2}}\phi_{3-j}
\label{2NPSE-SOC}
\end{eqnarray}
where $\phi_{j}(x,y,t)$ is a complex wave-function, with
normalization $\int_{-\infty}^{\infty}dx\, 
dy|\phi_{j}(x,y,t)|^2=N_{j}$, and
the conserved total number of atoms is $N=N_{1}+N_{2}$.
The corresponding axial widths  $\xi_{j}(x,t)$, are
\begin{eqnarray}
\xi_{j}^4 &=&
1+ \sqrt{2\pi} g_{jj} \xi_{j} \left|\phi_{j}\right|^2
+ 4\sqrt{\pi}g_{12} \frac{\xi_{j}^{4}}{(\xi^2_{1} +
\xi^2_{2})^{3/2}}\left|\phi_{3-j}\right|^2
\nonumber \\
&+& \sqrt{2}\Gamma (-1)^{j-1} \xi_{j}^{5/2} \xi^{1/2}_{3-j}
\frac{\xi_{1}^2 - \xi_{2}^2}{(\xi_{1}^2 + \xi_{2}^2)^{3/2}}
\Bigg(\frac { \phi_{1}^{*} \phi_{2} + \phi_{2}^{*}\phi_{1}}
{\left|\phi_{j}\right|^2 }\Bigg)
\label{widths-2D-mean-appox}
\end{eqnarray}
The system of coupled 2D-nonpolynomial Schr\"odinger
equations (\ref{2NPSE-SOC}), along with the equations 
(\ref{widths-2D-mean-appox}) is derived for the first time 
in this work.
In the symmetric case, when the strengths of the nonlinear 
interactions between different atomic states are equal,
$g_{11}=g_{22}=g_{12}\equiv g$, we have 
$\xi_{1}^4=\xi_{2}^4 \equiv \xi^4$.
The Eqs. (\ref{widths-2D-mean-appox}) taking the form
$\xi^4= 1+ \sqrt{2\pi} g\xi(\left|\phi_{1}\right|^2 + 
\left|\phi_{2}\right|^2)$. 
A binary BEC is self-attractive while $g<0$, and it is 
self-repulsive when $g>0$. 
This equation has exact 
solutions given by Cardano formula
\cite{Salasnich-2D, Salasnich-Malomed-NPSE-2D}.
In analogy with the 1D case, stationary solutions may be 
sought as 
$\phi_{j}(x,y,t)\rightarrow \phi_{j}(x,y)\exp{(-i\mu t)}$
with the restriction $\phi_{1}^*(x,y)=\phi_{2}(x,y)$. 
These stationary solutions were studied in
\cite {Salasnich-2D}.
The model (\ref{2NPSE-SOC}) is strictly two-dimensional 
under the condition $\xi_{j}^4= 1$ in the system of 
equations (\ref{widths-2D-mean-appox}).
By omitting the constant contribution of the axial energy 
$1/2$ (in units of $\omega_{z}$), we get the 2D GPE with SO
and Rabi couplings
\begin{eqnarray} 
{\mbox i} {\partial \over \partial t} \phi_{j} &=&
\Bigg [ -{\frac{1}{2}} \Bigg (\frac{\partial^2}{\partial x^2}
+\frac{\partial^2}{\partial y^2} \Bigg)
+ V(x,y) + (-1)^{j-1}{\mbox i}
\gamma \frac{\partial}{\partial x} 
\nonumber \\
&+& \sqrt{2\pi}g_{jj} 
\left| \phi_{j} \right|^2 +
\sqrt{2\pi}g_{12}
|\phi_{3-j}|^2 \Bigg] \phi_{j}+ \Gamma \phi_{3-j}
\label{2SE-SOC}
\end{eqnarray}

\section{Numerical results}
\label{IV}
We study the ground state of SO coupling BECs by solving
numerically the stationary 1D NPSE (\ref{Stationary-NPSE}) 
and the stationary 1D GPE obtained from (\ref{1D-eq})
for $g_{jj}=g_{12}\equiv g$, along with the condition 
$\phi_{1}^*(x)=\phi_{2}(x)$ and setting
$\Phi(x)\equiv \sqrt{2/N}\phi_{1}(x)$.
The numerical results are achieved using a split-step
Crank-Nicolson method with imaginary time propagation
\cite{Adhikari-numeric}.
The imaginary time propagation $(t\rightarrow-it)$ does not
preserve the normalization, because the time evolution 
operator is not unitary. 
To fix the normalization of the
wave function we need to includes the restoration of the 
normalization after each operation of Crank-Nicolson method,
just as in a BEC with a single component
\cite{Adhikari-numeric}.
This procedure is possible because the model given in 
equation (\ref{Stationary-NPSE})
represents a single 1D NPSE.
In a more general case, the restoration of the 
normalization of the two equations given by 
the generalized 1D NPSE (\ref{NPSE-SO-Rabi})
could be implemented according the
approaches outlined in \cite{normalization}.
Thus, we split the stationary 1D NPSE 
(\ref{Stationary-NPSE}), such that the derivative terms
(provided by the kinetic energy and the SO term),
the contribution of HO potential and the nonpolynomial
nonlinearity were  discretized and implemented as it 
has been presented in \cite{Adhikari-numeric}. 
The term associated with the Rabi coupling 
was handled by means of the adaptation of 
the numerical method implemented to obtaining
the ground state of atomic-molecular BECs
\cite{molecular-BEC-numeric}.
In a similar way it was solved the 1D GPE through
of the respective modification of the nonpolynomial 
contribution.
The spatial and time steps employed in the present work are
$\Delta x=0.01$ and $\Delta t=0.00025$, respectively.
The input wave-function is the known normalized 
Gaussian $\Phi=\pi^{-1/4}\exp(-x^{2}/2)$.

\begin{figure}[t] 
\begin{center}
\vskip .5cm
\includegraphics[width=10cm,clip]{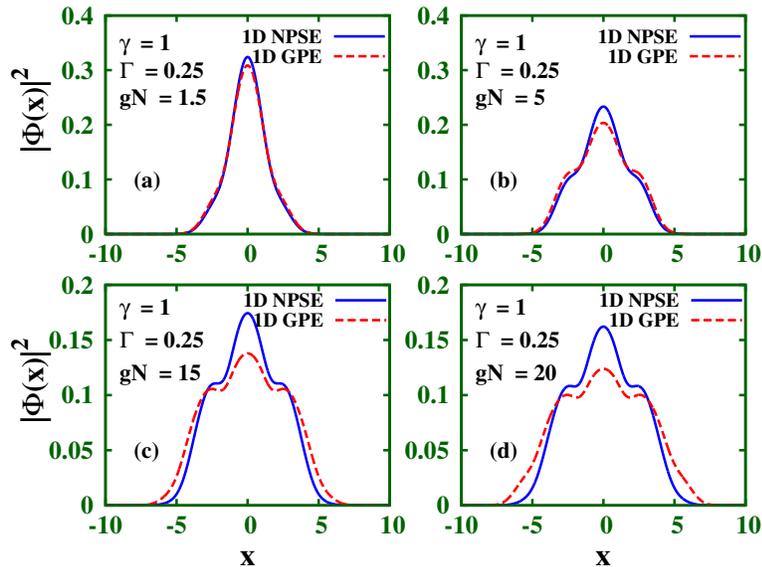}
\end{center}
\vskip -0.5cm
\caption{(Color online)
Density of a self-repulsive SO and Rabi coupled 
BEC under axial harmonic confinement given by 
$V(x)=x^{2}/18$. We set the SO coupling $\gamma=1$, 
the Rabi coupling $\Gamma=0.25$, and four values of the 
nonlinearity strength $gN$, panels (a)-(d). 
The solid line depicts the density predicted by the 
1D NPSE (\ref{Stationary-NPSE}) and
the dashed line represents the prediction of the stationary
1D GPE obtained by starting from (\ref{1D-eq}). 
Lengths are measured in units of the transverse confinement
radius $a_{\perp}=\sqrt{\hbar/(m\omega_{\perp})}$.}
\label{repulsive}
\end{figure}

We start the study of the ground state structure of the 
repulsive $(g>0)$ SO and Rabi coupled binary BEC in the
presence of a HO axial trap with frequency $\omega_x$, 
and anisotropy $\lambda\equiv\omega_{x}/\omega_{\perp}$,
such that $V(x)=(\lambda^{2}/2)x^2$. 
In Figure \ref{repulsive} (a)-(d) we plot the numerical 
results of the density profile
$|\Phi(x)|^2$ as a function of axial coordinate $x$.
We use $\lambda=1/3$, the SO coupling $\gamma=1$,
the Rabi coupling $\Gamma=0.25$, 
and four values of the nonlinearity strength $gN$.
Our findings show that for the constant values of $\gamma$ 
and $\Gamma$ and increasing the repulsive nonlinearity, the 
density profile expands and it decreases its height
displaying several local maxima (see the right top and the 
bottom panels in Figure \ref{repulsive}).
We compare the results of 1D NPSE (\ref{Stationary-NPSE})
with the stationary 1D GPE obtained starting from the model
(\ref{1D-eq}).
The density reduction with the increase of $gN$ 
is faster in the 1D GPE respect to the 1D NPSE.
In the weakly nonlinear regime $gN|\Phi(x)|^2\ll 1$
both models match very good each other. 
This fact can be verified in 
the Table \ref{tab:table1} with the values
of $gN|\Phi(x=0)|^2$.\\

\begin{table}[t]
\caption{\label{tab:table1} Values of $gN|\Phi(x=0)|^2$ in
a self-repulsive binary BEC with SO and Rabi couplings for
the 1D NPSE and the 1D GPE.}
\begin{indented}
\lineup
\item[]\begin{tabular}{@{}*{7}{l}}
\br
$gN$&\mbox{1.5}&\mbox{5}&\mbox{15}&\mbox{20}\cr
\mr
$gN|\Phi(x=0)|^2_{\,\,\,\,NPSE}$&\mbox{0.486}&
\mbox{1.167}&\mbox{2.615}
&\mbox{3.243}\cr
$gN|\Phi(x=0)|^2_{\,\,\,\,1D\,\,GPE}$&\mbox{0.463}&
\mbox{1.018}&\mbox{2.068} &\mbox{2.477}\cr
\br
\end{tabular}
\end{indented}
\vskip -.5cm
\end{table}

\begin{figure}[b] 
\begin{center}
\vskip .5cm
\includegraphics[width=10cm,clip]{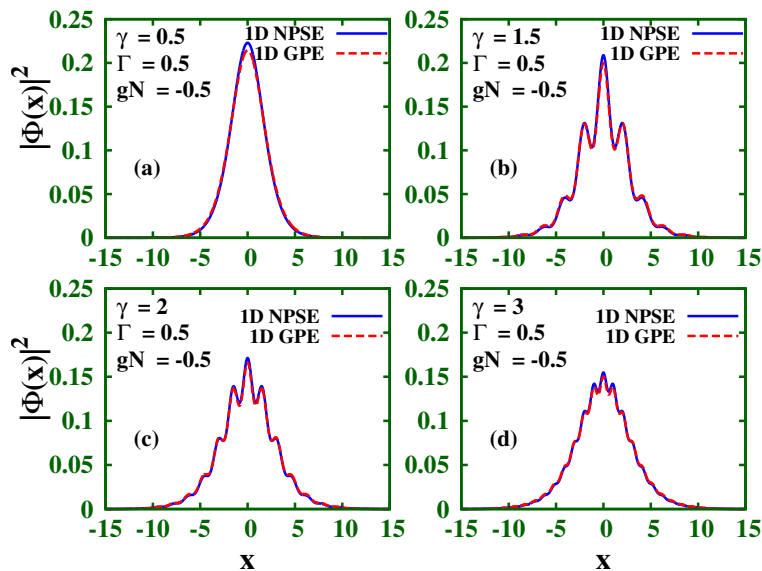}
\end{center}
\vskip -0.5cm
\caption{(Color online)
Density profile of a self-attractive SO and
Rabi coupled BEC without axial confinement, $V(x)=0$. 
We use the nonlinearity strength $gN=-0.5$, 
the Rabi coupling $\Gamma=0.25$, and four values of the
SO coupling $\gamma$, panels (a)-(d). 
The solid line is the solution given by the 1D NPSE
(\ref{Stationary-NPSE}), and the dashed line represents the 
prediction of the stationary 1D GPE given from (\ref{1D-eq}). 
Lengths are measured in units of the transverse confinement 
radius $a_{\perp}=\sqrt{\hbar/(m\omega_{\perp})}$.}
\label{attractive}
\end{figure}

\begin{figure}[t] 
\begin{center}
\vskip .5cm
\includegraphics[width=8.5cm,clip]{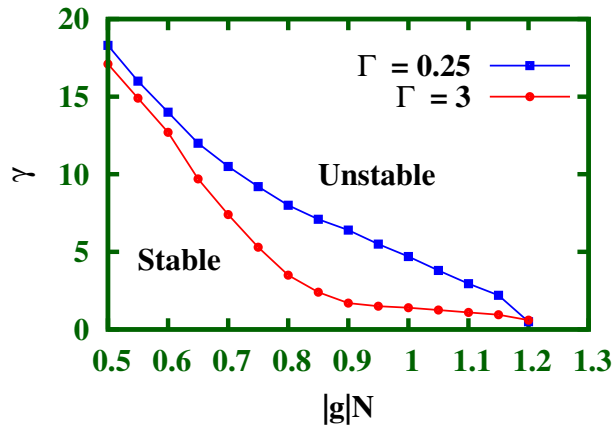}
\end{center}
\vskip -0.5cm
\caption{(Color online) Stability diagram of bright solitons in the SO and
Rabi coupled BEC. The SO coupling $\gamma$ versus the
nonlinearity strength $|g|N$ for two different values of
the Rabi coupling $\Gamma$, in dimensionless form.}
\label{unstable}
\end{figure}

It is also relevant analyzing the self-attractive $(g<0)$
SO and Rabi coupled BEC. Now, our aim is investigate 
effects of $gN$ on the bright solitons. 
In order to obtaining self-focusing nonlinear waves,
we solve numerically the 1D NPSE (\ref{Stationary-NPSE})
and the stationary 1D GPE obtained from (\ref{1D-eq}) 
with $V(x)=0$ and $g<0$.
In Figure \ref{attractive} (a)-(d) 
we plot the density profile
$|\Phi(x)|^2$ as a function of axial coordinate $x$
for the bright solitons, with $gN=-0.5$,
the Rabi coupling $\Gamma=0.5$, and four values of
SO coupling $\gamma$. Interesting issues are the change
of the shape of the soliton and the decreasing in the density
with the increasing of $\gamma$.
The results demonstrate that the interplay of the SO and 
Rabi couplings produces conspicuous sidelobes on the density 
profile of the bright solitons.
Salasnich and co-workers \cite{Bright-4}, demonstrate
an increase of soliton's height and the compression of
density profile for a constant $\gamma$ value and the 
increase of $\Gamma$.
For a constant value of Rabi coupling $\Gamma$ 
and the increasing of the SO coupling $\gamma$,
we shown that it is possible to obtaining a decrease 
in the density profile without a compression of the soliton. 
The agreement between the 1D NPSE and the 1D GPE is 
extremely good, and both models predict the existence of 
self-trapped matter-wave solitons 
as it is shown in the overlapping of the plots in 
Figure \ref{attractive}. This can be understood since 
the 1D NPSE with $gN = -0.5$ becomes effectively 1D.
However, it is worth noting, contrary to the 1D GPE, 
the 1D NPSE predicts the instability by collapse.
Now, we analyze some aspects of the collapse in the SO and
Rabi couplings BECs by means of the 1D NPSE.
It is known that for a self-attractive condensate 
without SO-coupling and/or with Rabi coupling, 
the model (\ref{Stationary-NPSE}) with $V(x)=0$ predicts
the existence of bright solitons only for
$-4/3<gN<0$ \cite{vector-solit}.
By solving the Eq. (\ref{Stationary-NPSE}) for 
bright solitons in the absence of the axial confinement,
we plot in Figure \ref{unstable} the positive SO coupling 
$(\gamma>0)$ as function of the strength
$|g|N$ for two different values of $\Gamma$.
Similar plots exist in the range $0.25<\Gamma<3$.
With the increasing of the attraction the density 
profile of the soliton shrinks, and once the strength 
of the interaction exceeds the critical value of $|g|N$,
the condensate becomes so attractive that growth peak
height is stopped by inelastic collisions and the system
must collapse. We find the critical value $|g|N\simeq 1.2$ 
beyond which the condensate collapses, and it is not 
dependent on the Rabi coupling.
We verified a reduction of the collapse threshold
for bright solitons under the action of the SO and Rabi
couplings, initially predicted in \cite{Bright-4}.
The 1D NPSE (\ref{Stationary-NPSE})  
gives rise to the collapse of the condensate,
when the density of the condensate in the denominator 
of the nonlinear term reaches a critical value 
$|\Phi|^{2}=(|g|N)^{-1}$
\cite{Bright-4,Salasnich-NPSE}.
This implies that the effective nonpolynomial
nonlinearity in the 1D NPSE
is quite essential compared with the cubic nonlinearity in 
1D GPE where there is no collapse.
The reduction in the SO coupling caused by the increasing of
the strength of the interaction is reflected in the density 
of the soliton as an increasing in the real component of 
the wave function and decreasing in the imaginary one, 
together with decreasing in the oscillations in both parts
of the wave function.
Now we consider the possible origin of the unstable
region in Figure \ref{unstable}.
To achieve this purpose we fix the values of Rabi coupling 
$\Gamma$ and the strength of interaction $|g|N$,
and we increase the SO coupling $\gamma$. 
Above a critical value determined by the plots in Figure
\ref{unstable}, the oscillations set by the SO coupling 
becomes unstable the condensate and it is destroyed.
This unstable region could be experimentally analyzed 
by means of a recently technique implemented in 
\cite{Tunable-SOC}. 
Here it has been proposed a scheme for controlling 
SO-coupling between two hyperfine ground states 
in a binary BEC through a fast and coherent 
modulation of the Raman laser intensities.
The above results also apply at $\gamma<0$. In that case 
we have $x\rightarrow -x$ in 1D NPSE (\ref{Stationary-NPSE})
and the diagram of Figure \ref{unstable} is symmetric.

\section{Summary and outlook}
\label{V} 
We have analyzed the 1D and 2D dimensional reduction of a
binary
bosonic quantum field theory in which the two components 
are coupled by the nonlinear inter-atomic interactions
and the SO and Rabi couplings.
We have derived for the first time, an 
effective 1D and 2D quantum Hamiltonian with the 
corresponding effective 1D and 2D nonpolynomial Heisenberg 
equations, for both self-repulsive and self-attractive 
inter-atomic interactions.
These 1D and 2D models open the door to a new level
in the research of interacting gases of bosons with SO and 
Rabi couplings at finite temperature in reduced dimensions.
Recently, dramatic implications were predicted about the
effect of the SO coupling on thermal properties of
bosonic gases \cite{T-finite}, 
showing thus the amplification of quantum
fluctuations induced by the SO coupling.
We show the possibility of obtaining two models in a
mean-field approximation derived from the quantum field 
equations in the case in which the many-body quantum state 
of the system coincides with the Glauber coherent state.
Our findings are agree with the system of coupled NPSEs 
considered in the study of dynamics of BECs with SO and
Rabi couplings in one \cite {Bright-4} and two dimensions
\cite{Salasnich-2D}.
The numerical results demonstrate the emergence of
conspicuous sidelobes on the density profile due to the 
interplay of SO and Rabi couplings, for both signs of the 
interaction.
The mean-field approximation given by 1D NPSE in repuslive
BECs with SO and Rabi couplings shows the relevance of 
nonlinearity compared to 1D GPE.
An interesting result in the 1D NPSE describing self-attractive 
binary BECs, is the possibility of getting an unstable
condensate by fixing the Rabi coupling and the strength of
interaction and by increasing of SO coupling.
Thus showing a new instability in addition to the
existing collapse threshold, 
established by the strength of interaction.

The effective 1D and 2D models can be used to study the 
effect of inhomogeneous nonlinearity, as well as general
forms of the couplings.
One may expect interesting issues with non trivial effects
by extending the present analysis for matter-wave
vortices. In the absence of two couplings, 
vortical states in a BEC have been studied by means of 
1D \cite{Vortices-1D-NPSE} and
2D \cite{Salasnich-Malomed-NPSE-2D} NPSEs.

\newcommand{\noopsort}[1]{} \newcommand{\printfirst}[2]{#1}
\newcommand{\singleletter}[1]{#1} \newcommand{\switchargs}[2]{#2#1}


\section*{References}
\begin{thebibliography}{99}

\bibitem{SOC-1} Lin Y-J, Jim\'enez-Garc\'ia K, and 
Spielman I B 2011 Nature (London) {\bf 471}, 83

\bibitem {soc-1} Zhang J-Y, Ji S-C, Chen Z, Zhang L,
Du Z-D, Yan B, Pan G-S , Zhao B, Deng Y-J, Zhai H, Chen
S, and J-W Pan 2012 Phys. Rev. Lett. {\bf 109}, 115301

\bibitem {SOC-2} Wang P, Yu Z Q, Fu Z, Miao J, Huang L,
Chai S, Zhai H, and Zhang J 2012
Phys. Rev. Lett. {\bf 109}, 095301

\bibitem {SOC-3} Cheuk L W, Sommer A T, Hadzibabic Z, 
Yefsah T, Bakr W S, and Zwierlein M W 2012
Phys. Rev. Lett. {\bf 109}, 095302

\bibitem{spin-orbit} Zhai H 2012 
Int. J. Mod. Phys. B {\bf 26}, 1230001

\bibitem{spin-orbit-1} Zheng W, Yu Z-Q, Cui X and Zhai H
2013 J. Phys. B: At. Mol. Opt. Phys. {\bf46}, 134007

\bibitem {vort-1} Ho T L and Zhang S Z 2011
Phys. Rev. Lett. {\bf 107}, 150403

\bibitem {vort-2} Xu X-Q and Han J H 2011
Phys. Rev. Lett. {\bf 107}, 200401

\bibitem {vort-3} Radic J, Sedrakyan T A, Spielman I B,
and Galitski V 2011 Phys. Rev. A {\bf 84}, 063604

\bibitem {vort-4} Zhou X F, Zhou J, and Wu C J 2011
Phys. Rev. A {\bf 84}, 063624

\bibitem {Santos-1} Sinha S, Nath R, and Santos L 2011
Phys. Rev. Lett. {\bf 107}, 270401

\bibitem{vortex-5} Hu H, Ramachandhran B, Pu H, and
Liu X-J 2012 Phys. Rev. Lett. {\bf108}, 010402 

\bibitem{half-vortex} Ramachandhran B, Opanchuk B,
Liu X-J, Pu H, Drummond P D, and Hu H 2012
Phys. Rev. A {\bf 85}, 023606

\bibitem{spin1} Wen L, Sun Q, Wang H Q, Ji A C, and 
Liu W M 2012 Phys. Rev. A {\bf86}, 043602

\bibitem{Spinor-BEC} Wang C, Gao C, Jian C M,
and Zhai H 2010 Phys. Rev. Lett.  {\bf 105}, 160403


\bibitem{SOC-BECs} Ruokokoski E, Huhtam\"aki J A M, and
M\"ott\"onen M 2012 Phys. Rev. A {\bf 86}, 051607(R)

\bibitem{Tricriticality} Li Y, Pitaevskii  L P,
and Stringari S 2012 Phys. Rev. Lett. {\bf 108}, 225301

\bibitem{order-disorder} Barnett R, Powell S,
Graβ T, Lewenstein M, and Das Sarma S 2012
Phys. Rev. A {\bf85}, 023615

\bibitem{Malomed-1} Zezyulin D A, Driben R, Konotop V V,
and Malomed B A 2013 Phys. Rev. A {\bf 88}, 013607

\bibitem{SOC-experim} Deng Y, Cheng J, Jing H, Sun C-P, 
and Yi S 2012 Phys. Rev. Lett. {\bf 108}, 125301

\bibitem{Bright-1} Xu Y, Zhang Y, and Wu B 2013
Phys. Rev. A {\bf87}, 013614

\bibitem{Bright-2} Achilleos V, Frantzeskakis D J,
Kevrekidis P G, and Pelinovsky D E 2013
Phys. Rev. Lett. {\bf 110}, 264101

\bibitem{Bright-3} Kartashov Y V, Konotop V V, and
Abdullaev F Kh 2013 Phys. Rev. Lett. {\bf 111}, 060402

\bibitem{Bright-4} Salasnich L and Malomed B A 2013
Phys. Rev. A {\bf 87}, 063625

\bibitem{2D-1} Sakaguchi H, Li B, and Malomed B A 2014
Phys. Rev. E {\bf 89}, 032920

\bibitem{2D-2} Lobanov V E, Kartashov  Y V,
and Konotop V V 2014 Phys. Rev. Lett. {\bf 112}, 180403

\bibitem{Salasnich-2D} Salasnich L, Cardoso  W B, and
Malomed B A 2014 Phys. Rev. A {\bf 90}, 033629

\bibitem{Vortex-dynamics} Sakaguchi H. and Li B 2013
Phys. Rev. A {\bf 87}, 015602 

\bibitem{Fetter} Fetter A L 2014 Phys. Rev. A {\bf 89},
023629

\bibitem{Fetter-2} Fetter A L 2015 J. Low Temp. Phys.
{\bf 180}, 37

\bibitem{Adhikari-1} Cheng  Y S, Tang  G H, and 
Adhikari S K 2014 Phys. Rev. A {\bf 89}, 063602

\bibitem{stability-1} He P-S 2013 Eur. Phys. J. D {\bf 67}, 
48

\bibitem{stability-2} He P-S, You W-L, and Liu W-M 2013
Phys. Rev. A {\bf 87}, 063603

\bibitem{Adhikari-2} Gautam S and Adhikari S K 2015
Laser Phys. Lett. {\bf 12} 045501

\bibitem{Adhikari-3}  Gautam S and Adhikari S K 2015 
Phys. Rev. A {\bf 91}  063617

\bibitem{Collapse-SOC-BEC} Mardonov Sh, Sherman E Ya,
Muga J G, Wang Hong-Wei, Ban Yue, and Chen Xi 2015
Phys. Rev. A {\bf 91}, 043604

\bibitem{Josephson-1} Zhang D-W, Fu L-B, Wang Z D,
and Zhu S-L 2012 Phys. Rev. A {\bf 85}, 043609

\bibitem{Josephson-2} Garcia-March M A, Mazzarella G,  
Dell'Anna L, Juli\'a-D\'iaz B, Salasnich L, and Polls A
2014 Phys. Rev. A {\bf 89}, 063607

\bibitem{Bright-solitons1} Cornish S L, Thompson S T, 
and Wieman C E 2006 Phys. Rev. Lett. {\bf 96}, 170401

\bibitem{Dalfovo} Dalfovo F, Giorgini S, Pitaevskii L P, 
and Stringari S 1999 Rev. Mod. Phys. {\bf 71} 463

\bibitem{Bright-solitons2} Eiermann B, Anker Th, Albiez M,
Taglieber M, Treutlein P, Marzlin K-P, and Oberthaler M K 
2004 Phys. Rev. Lett. {\bf 92}, 230401 

\bibitem{Giamarchi} Giamarchi T 2004
\textit{Quantum Physics in One Dimension
(Great Britain: Oxford University Press)}

\bibitem{cazalilla2011} Cazalilla M A, Citro R, 
Giamarchi T, Orignac E, and Rigol M 2011
Rev. Mod Phys. {\bf 83} 1405

\bibitem{Muñoz-Delgado} Mu\~noz Mateo A, and Delgado V
2008 Phys. Rev. A {\bf 77}, 013617

\bibitem{Salasnich-NPSE} Salasnich L, Parola A, and
Reatto L 2002 Phys. Rev. A {\bf 65}, 043614

\bibitem{Dim-red-dip} Sinha S and Santos L
2007 Phys. Rev. Lett. {\bf 99}, 140406 

Muruganandam P and Adhikari S K 2012 
Laser Phys. {\bf 22}, 813-20

Chiquillo Emerson 2014 Laser Phys. {\bf 24}, 
085502

\bibitem{T-finite} Liao R, Huang Z-G, Lin X-M, 
and Fialko O 2014 Phys. Rev. A {\bf 89}, 063614

\bibitem{Salasnich-Quantum-solitons} Barbiero L and 
Salasnich L 2014 Phys. Rev. A {\bf 89}, 063605

\bibitem{Salasnich-QFT} Salasnich L 2015
\textit{Quodons in Mica (Springer series in materials 
science vol 221
Discrete bright solitons in Bose-Einstein condensates
and dimensional reduction in quantum field theory)}
ed Archilla J F R , Jim\'enez N, S\'anchez-Morcillo V J, and
Garc\'ia-Raffi L M (Springer International Publishing)

\bibitem{Salasnich-book} Salasnich L 2014
\textit{Quantum Physics of Light and Matter.  
A Modern Introduction to Photons, Atoms and Many-Body 
Systems (Cham: Springer)} p. 167


\bibitem{Coherent-states-QFT} Stoof H T C, Gubbels K,
Dickerscheid D 2009 \textit{Ultracold Quantum Fields
(Berlin: Springer)} p. 131

\bibitem{Coherent-states-1} 
Rogel-Salazar J, Choi S, New G H C, and Burnett K 2004
J. Opt. B: Quantum Semiclass. Opt. {\bf 6} R33-R59

\bibitem{Coherent-states-2} Glauber R 1963
Phys. Rev. {\bf 131}, 2766


Zhang W-M, Feng D H, 
and Gilmore R 1990 Rev. Mod. Phys. {\bf 62}, 867

\bibitem{vector-solit} Salasnich L. and Malomed B A 2006
Phys. Rev. A {\bf 74}, 053610

\bibitem{GPE-SOC-1D} Achilleos V, Stockhofe J,
Kevrekidis P G, Frantzeskakis D J, and Schmelcher P 2013
Eur. Phys. Lett. {\bf 103},  20002

\bibitem{Salasnich-Malomed-NPSE-2D} Salasnich L and 
Malomed B A 2009 Phys. Rev. A {\bf 79}, 053620

\bibitem{Adhikari-numeric}
Muruganandam P and Adhikari S K 2009 Comput. Phys.
Commun. {\bf 180}, 1888 

\bibitem{normalization} Bao W and Cai Y 2015
Siam J. Appl. Math. {\bf 75}, 492

Bao W and Cai Y 2011 
East Asia Journal on Applied Mathematics {\bf 1}, 49-81

\bibitem{molecular-BEC-numeric} Jiang W, Wang H, and
Li X 2013 Comput. Phys. Commun. {\bf 184}, 2396

\bibitem{Tunable-SOC} 
Zhang Y, Chen G, and Zhang C 2013
Sci. Rep. {\bf 3}, 1937

Lian J, Yu L, Liang J-Q, Chen G, and Jia S 2013
Sci. Rep. {\bf 3} 3166 

Jim\'enez-Garc\'ia K, LeBlanc L J, Williams R A,
Beeler M C, Qu C, Gong M, Zhang C, and Spielman I B 
2015 Phys. Rev. Lett. {\bf 114}, 125301

\bibitem{Vortices-1D-NPSE} Salasnich L, Malomed B A, 
and Toigo F 2007 Phys. Rev. A {\bf 76}, 063614



\end{thebibliography}
\end{document}